\documentclass[sigconf]{acmart}
\usepackage{enumitem}
\usepackage{booktabs}
\usepackage{tabularx}
\usepackage{refcount}
\usepackage{makecell}
\usepackage[normalem]{ulem}
\usepackage{multirow}
\usepackage[toc,page]{appendix}
\usepackage{pifont}
\usepackage[cjk]{kotex}
\usepackage{array}
\newcommand{\tabitem}{~~\llap{\textbullet}~~}
\useunder{\uline}{\ul}{}

\AtBeginDocument{%
  \providecommand\BibTeX{{%
    \normalfont B\kern-0.5em{\scshape i\kern-0.25em b}\kern-0.8em\TeX}}}

\copyrightyear{2023}
\acmYear{2023}
\setcopyright{acmlicensed}
\acmConference[L@S '23] {Proceedings of the Tenth ACM Conference on Learning @ Scale}{July 20--22, 2023}{Copenhagen, Denmark.}
\acmBooktitle{Proceedings of the Tenth ACM Conference on Learning @ Scale (L@S '23), July 20--22, 2023, Copenhagen, Denmark}
\acmPrice{15.00}
\acmISBN{979-8-4007-0025-5/23/07}
\acmDOI{10.1145/3573051.3596200}

\begin{document}


\title{RECIPE: How to Integrate ChatGPT into EFL Writing Education}


\author{Jieun Han}
\authornote{Authors contributed equally.}
\affiliation{
    \institution{KAIST}
    \state{Daejeon}
    \country{South Korea}
}
\email{jieun_han@kaist.ac.kr}

\author{Haneul Yoo}
\authornotemark[1]
\affiliation{
    \institution{KAIST}
    \state{Daejeon}
    \country{South Korea}
}
\email{haneul.yoo@kaist.ac.kr}

\author{Yoonsu Kim}
\affiliation{
    \institution{KAIST}
    \state{Daejeon}
    \country{South Korea}
}
\email{yoonsu16@kaist.ac.kr}

\author{Junho Myung}
\affiliation{
    \institution{KAIST}
    \state{Daejeon}
    \country{South Korea}
}
\email{junho00211@kaist.ac.kr}

\author{Minsun Kim}
\affiliation{
    \institution{KAIST}
    \state{Daejeon}
    \country{South Korea}
}
\email{9909cindy@kaist.ac.kr}

\author{Hyunseung Lim}
\affiliation{
    \institution{KAIST}
    \state{Daejeon}
    \country{South Korea}
}
\email{charlie9807@kaist.ac.kr}

\author{Juho Kim}
\affiliation{
    \institution{KAIST}
    \state{Daejeon}
    \country{South Korea}
}
\email{juhokim@kaist.ac.kr}

\author{Tak Yeon Lee}
\affiliation{
    \institution{KAIST}
    \state{Daejeon}
    \country{South Korea}
}
\email{takyeonlee@kaist.ac.kr}

\author{Hwajung Hong}
\affiliation{
    \institution{KAIST}
    \state{Daejeon}
    \country{South Korea}
}
\email{hwajung@kaist.ac.kr}

\author{So-Yeon Ahn}
\authornotemark[2]
\affiliation{
    \institution{KAIST}
    \state{Daejeon}
    \country{South Korea}
}
\email{ahnsoyeon@kaist.ac.kr}

\author{Alice Oh}
\authornote{Corresponding authors}
\affiliation{
    \institution{KAIST}
    \state{Daejeon}
    \country{South Korea}
}
\email{alice.oh@kaist.edu}

\renewcommand{\shortauthors}{Han et al.}

\begin{abstract}

The integration of generative AI in the field of education is actively being explored. In particular, ChatGPT has garnered significant interest, offering an opportunity to examine its effectiveness in English as a foreign language (EFL) education. To address this need, we present a novel learning platform called \texttt{RECIPE} (Revising an Essay with ChatGPT on an Interactive Platform for EFL learners). Our platform features two types of prompts that facilitate conversations between ChatGPT and students: (1) a hidden prompt for ChatGPT to take an EFL teacher role and (2) an open prompt for students to initiate a dialogue with a self-written summary of what they have learned. We deployed this platform for 213 undergraduate and graduate students enrolled in EFL writing courses and seven instructors. For this study, we collect students' interaction data from \texttt{RECIPE}, including students' perceptions and usage of the platform, and user scenarios are examined with the data. We also conduct a focus group interview with six students and an individual interview with one EFL instructor to explore design opportunities for leveraging generative AI models in the field of EFL education. 
\end{abstract}



\begin{teaserfigure}
  \includegraphics[width=.96\textwidth]{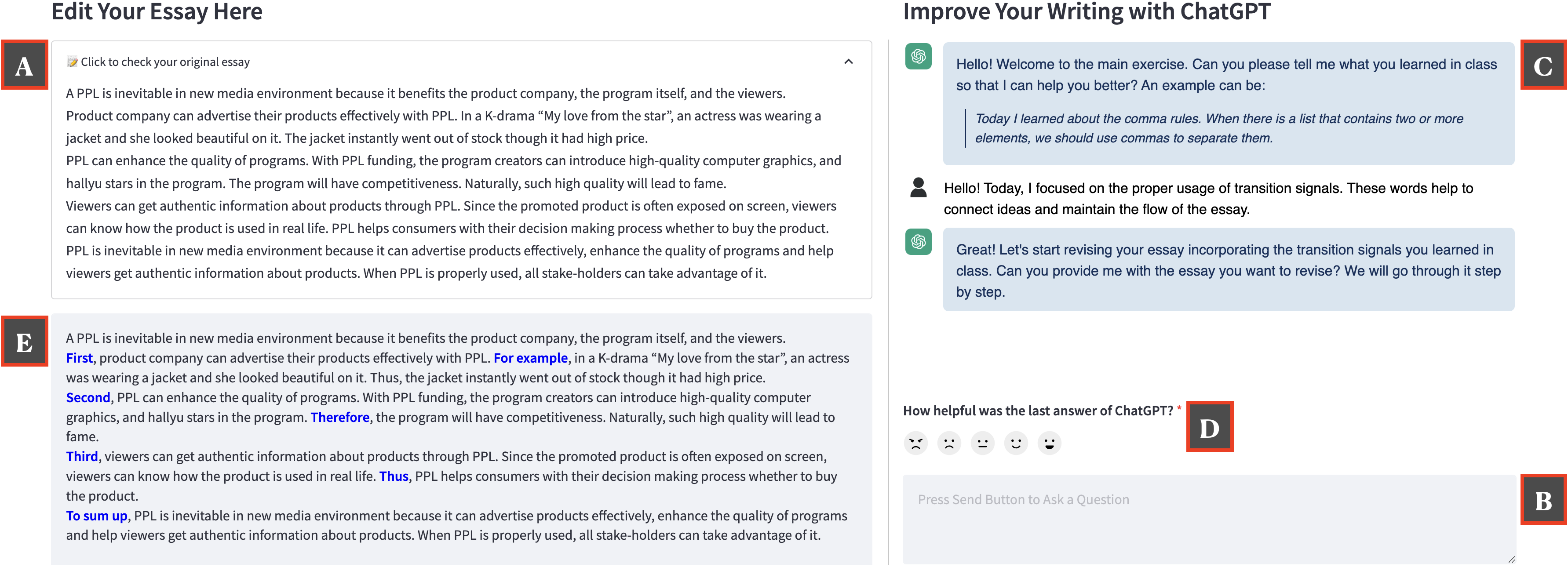}
  \caption{A screenshot of the writing exercise in our platform. A student's original essay is initially provided at (A). A student starts a conversation with ChatGPT at (B) by summarizing what they learned in the previous lecture as we instructed. Conversation history is logged at (C), and a student is prompted to assess the performance of ChatGPT's every response in a 5-point Likert scale at (D). Then, a student can revise the essay at (E) and repeat the process from (B) to (E).}
  \label{fig:main_exercise}
\end{teaserfigure}


\maketitle

\section{Introduction}
In the context of English as a foreign language (EFL) education, the integration of artificial intelligence (AI) technology has been shown to enhance students' learning experience. The use of AI-based tools, including Grammarly\thinspace\footnote{\label{grammarly}\url{https://www.grammarly.com/}} and Quillbots\thinspace\footnote{\url{https://quillbot.com/}}, has resulted in significant improvement in EFL learners' writing abilities. Moreover, learners have exhibited positive perceptions towards the use of these tools in their writing class \cite{huang2020the, ariyanti2021technology}.

ChatGPT\thinspace\footnote{\label{chatgpt}\url{https://chat.openai.com/}}, a large language model (LLM)-driven chatbot by OpenAI, has made a significant breakthrough in the domain of language learning. While earlier chatbots were unnatural and incapable of engaging language learners \cite{huang2022chatbots}, ChatGPT can generate natural and personalized responses, which make students' learning experience more interactive and engaging \cite{sullivan2023chatgpt, rudolph2023chatgpt}. Recent studies have suggested the potential educational benefits of incorporating generative AI into education \cite{milano2023large}. Hence many online educational platforms such as Khan Academy\thinspace\footnote{\url{https://www.khanacademy.org/}} and Duolingo\thinspace\footnote{\url{https://www.duolingo.com/}} have already started integrating ChatGPT to their functionalities. 

Despite such suggestions and attempts to incorporate ChatGPT into language education, only episodic and anecdotal knowledge has been shared rather than systematic investigation \cite {qadir2022engineering, baidoo2023education}. Therefore, we need to examine the effective use of ChatGPT in higher education and identify its learning effect. In order to effectively integrate ChatGPT into EFL writing courses, it is crucial to investigate the specific design of the integration, considering learners' perceptions and usage of ChatGPT in higher education. In this regard, we introduce an interactive learning platform, \texttt{RECIPE} (Revising an Essay with ChatGPT on an Interactive Platform with EFL learners), aiming to leverage the data collected from our platform to guide learners towards a more effective and engaging learning experience. 
The main contributions of this work are as follows:
\begin{enumerate}
    \item We analyze the perception and usage of ChatGPT in the context of EFL among learners and instructors.
    \item We introduce \texttt{RECIPE} (Revising an Essay with ChatGPT on an Interactive Platform with EFL learners), a learning platform designed to integrate ChatGPT with underlying two types of prompts for students' better learning experience.
    \item We collect students' interaction data on their perception and ChatGPT usage through \texttt{RECIPE} to investigate further development of our platform.
    
\end{enumerate}

\section{Preliminary Questionnaire}
In order to investigate students' attitudes, usage, and expectations of ChatGPT, especially in the context of EFL, we conducted a preliminary questionnaire with 213 (91 undergraduate and 122 graduate) college students in South Korea. They were enrolled in three types of English writing courses for the spring semester of 2023. In specific, 91 undergraduate students were enrolled either the Intermediate English Reading \& Writing (IRW; 46 enrolled) or the Advanced English Writing (AW; 45 enrolled) course, based on their TOEFL writing scores (15-18 for IRW, and 19-21 for AW). 122 graduate students were enrolled in Scientific Writing (SW).

The questionnaire consists of 5-point Likert scale questions about participant's prior experience and evaluation of ChatGPT, with regard to \textsc{perceived helpfulness}, \textsc{trustworthiness}, \textsc{credibility}, \textsc{appropriateness of style/tone}, \textsc{performance}, \textsc{overall satisfaction}, and \textsc{referral intention}\thinspace\footnote{to what extent they would recommend others to adopt ChatGPT} of ChatGPT in the context of college education.
Detailed questions are described in Appendix \ref{appendix:preliminary_study}.
Results from this survey provide visions for developing an educational platform with ChatGPT. 

\subsection{Perception and Usage of ChatGPT}
The majority of students reported a positive user experience with ChatGPT across all seven factors described above.
Specifically, 85\% of students with ChatGPT exposure used the tool in their academic work, indicating a potential for leveraging and developing the technology for educational purposes.
Moreover, less than half of the students used ChatGPT to improve their English writing skills, highlighting an opportunity and a need for research on integrating ChatGPT into EFL education. 
These students also relied on other widely available AI tools including Grammarly\footnotemark[\getrefnumber{grammarly}], Turnitin\thinspace\footnote{\url{https://www.turnitin.com/}}, and Google Translate\thinspace\footnote{\url{https://translate.google.com/}}, Papago\thinspace\footnote{\url{https://papago.naver.com/}}, Wordtune\thinspace\footnote{\url{https://www.wordtune.com/}}, ExplainPaper\thinspace\footnote{\url{https://www.explainpaper.com/}}, and Elicit\thinspace\footnote{\url{https://elicit.org/}} for help with English writing (90.4\%) most, followed by reading (61.6\%), grammar (57.1\%), speaking (27.8\%), and listening (12.6\%) skills.

This questionnaire also underscores the need for comprehensive and well-guided instructions for effectively and efficiently integrating ChatGPT into EFL writing courses, rather than implementing an LLM-agnostic class policy and allowing students to use ChatGPT on their own discretion.

Students with prior experience and knowledge of LLMs expressed higher levels of satisfaction and expectations of using ChatGPT in the courses they enrolled. 
In particular, those who had low LLM understanding and enrolled in IRW and AW, expected that ChatGPT would be useful for asking questions about the lecture and finding sources to support their writing, which are notable limitations of ChatGPT with hallucination \cite{bang2023multitask}. Whereas students with a high level of LLM understanding exhibited greater satisfaction with their previous ChatGPT experience regarding \textsc{perceived helpfulness}, \textsc{appropriateness of style/tone}, \textsc{performance}, \textsc{overall satisfaction}, and \textsc{encouragement}, compared to those with low LLM understanding. However, we did not find a statistically significant difference in terms of \textsc{trustworthiness} and \textsc{credibility}.
Students with a high level of LLM understanding and those with ChatGPT exposure also exhibited significantly positive expectations regarding \textsc{performance}, \textsc{credibility}, and \textsc{overall satisfaction} towards the use of ChatGPT in academic courses.
Based on these findings, we posit that our proposed platform \texttt{RECIPE} can effectively and efficiently guide students towards obtaining satisfactory responses from ChatGPT, ultimately enhancing their English writing skills.

\section{Platform Design}
\begin{table*}[htb!]
    \caption{Data collected during each phase of the platform.}
    \label{tab:platform_data}
    \centering
    \begin{tabularx}{\textwidth}{@{}l|X|X|X@{}}
        \toprule
        \multicolumn{1}{c|}{}& \multicolumn{1}{c|}{Pre-survey}                                                                                     & \multicolumn{1}{c|}{Writing Exercise}                                                                               & \multicolumn{1}{c}{Post-survey}                                                                                                   \\ \midrule
        (a) Essay            & \makecell[tl]{\tabitem Students' original essay}                                                                    & \makecell[tl]{\tabitem Student's edited essay and its revision \\history}                                        & \makecell[tl]{\tabitem Reason not to revise essay\\ \tabitem Confidence on students' essay}                                        \\ \midrule
        (b) ChatGPT          & \makecell[tl]{\\}                                                                                                   & \makecell[tl]{\tabitem Conversation log data\\ \tabitem Helpfulness of each ChatGPT response} & \makecell[tl]{\tabitem Topics discussed with ChatGPT\\ \tabitem Overall satisfaction towards \\ChatGPT responses}                            \\ \midrule
        (c) Student          & \makecell[tl]{\tabitem Expected help for the session\\ \tabitem Comprehension level about course \\topics} & \makecell[tl]{\tabitem Timestamp of each sent message}                                                                                              & \makecell[tl]{\tabitem Actual help through the session\\ \tabitem Improvements in comprehension \\level after writing exercise} \\ \bottomrule
    \end{tabularx}
\end{table*}

This section outlines the design of \texttt{RECIPE}, geared towards the development of English writing skills, utilizing ChatGPT’s capabilities in a targeted
and effective manner. We deployed \texttt{RECIPE} for 213 students who enrolled to EFL courses (IRW, AW, and SW) in the spring semester of 2023. Students' interaction data is  gathered throughout the semester from undergraduate and graduate students using our platform.
\texttt{RECIPE} comprises three main components: a pre-survey (\S\ref{sec:pre-survey}), the writing exercise (\S\ref{sec:main_exercise}), and a post-survey (\S\ref{sec:post-survey}).
Table \ref{tab:platform_data} displays the data collected at each phase of the platform.  

\subsection{Pre-survey}\label{sec:pre-survey}

We designed a pre-survey to collect (1) students’ expectations for the upcoming exercise and (2) their understanding of the topics that they will discuss with ChatGPT during the exercise. The pre-survey aims to compare students' expected and actual assistance and to examine changes in their understanding before and after the exercise. 

Students are asked to indicate all applicable objectives that they expect to receive assistance with during the exercise by selecting from a list of nine options provided in Table \ref{tab:expectation_academic_purpose} of Appendix \ref{appendix:preliminary_study}.
Additionally, students are required to rate their understanding of the topics to be addressed during the exercise, using a 5-point Likert scale.

\subsection{Writing Exercise}\label{sec:main_exercise}

Figure \ref{fig:main_exercise} displays a screenshot of our platform, which has a writing exercise interface divided into two sections: the left side is for editing an essay, while the right side is for interacting with ChatGPT. The students’ essay from the previous session is available on the left side, enabling them to revise their work with the help of ChatGPT.

\begin{table*}[!htb]
    \caption{Initial prompts at writing exercise}
    \label{tab:initial_prompt}
    \centering
    \begin{tabularx}{\textwidth}{@{}X|X@{}}
    \toprule
    \multicolumn{1}{c|}{(a) Hidden prompt for ChatGPT}                                                                                                                                                                                                                                                                    & \multicolumn{1}{c}{(b) Open prompt for students}                                                                                                                                                                                                                                        \\ \midrule
    Act as an English writing class teacher and instruct a student to revise an essay based on the content learned during class. The student will provide you with a summary of what they learned. Please guide them step by step and do not revise or rewrite the essay at once. If you understand, please say yes only. & Hello! Welcome to the main exercise. Can you please tell me what you learned in class so that I can help you in a better way? An example can be: ``Today I learned about comma rules. When there is a list that contains two or more elements, we should use commas to separate them.'' \\ \bottomrule
    \end{tabularx}
\end{table*}

On the right side, students can consult ChatGPT about their essay.
Unlike the existing ChatGPT interface from OpenAI, our platform provides two initial prompts based on empirical prompt engineering, as shown in Table \ref{tab:initial_prompt}: (a) a hidden prompt for the model to set a persona for ChatGPT, acting as a personalized English writing teacher, and (b) an open prompt for students to start a dialogue efficiently.
ChatGPT instructs students step by step to revise their essay based on the content they learned during the class, and students are asked to summarize what they learned during the corresponding week or previous classes as the first dialogue.
We advise both ChatGPT and students not to provide or request a revised version of the entire essay without any explanation.
These instructions serve two purposes: first, to remind students of the lecture content and enhance their learning, and second, to help students receive a more class-relevant response from ChatGPT. 
We believe this suggested interface can guide EFL learners to write a more specific opening prompt for ChatGPT.

After each turn of the conversation, the edited version of the students' essay will be saved to analyze how they revised it based on ChatGPT's response. In addition, students are asked to rate the helpfulness of the response generated by ChatGPT using a 5-scale Likert scale. Students can continue to converse with ChatGPT and revise their essay.

\subsection{Post-survey}\label{sec:post-survey}


First, to gain insights into students' decision-making regarding essay revision, we inquire about their reasons for not making an edit. As the essay editing history is collected during the writing exercise, we can compare the original version of the student’s essay with the submitted version after the conversation with ChatGPT. 
To further analyze prompt engineering, we ask students about the main topic of their conversation using multi-select questions, including whether it was about the current week’s lecture, previous lectures, and content not covered in the lecture. 
To compare students’ expectations of the exercise with their actual usage of ChatGPT, we provide the same nine options mentioned in the pre-survey (\S\ref{sec:pre-survey}).
Additionally, we collect data on students’ confidence in their essay and overall satisfaction with ChatGPT’s responses to analyze their experiences throughout the semester.
Finally, students share how they utilized ChatGPT's responses to modify their essay and their experiences regarding it in a free-text form.

\subsection{User Scenario}\label{sec:user_scenario}

Undergraduate courses (IRW and AW) and graduate courses (SW) have different requirements, as SW is focused on writing scientific research papers while IRW and AW are intended for essay writing. In order to better align with these varying needs, we have designed separate user scenarios. 

For IRW and AW, we collect in-class timed essay data from students before they use our platform. Students are provided with a prompt by their instructors and asked to write an argumentative essay. The essay is written in class within 40 minutes with pens and pencils, ensuring that they cannot rely on online tools for assistance. The essays are digitized and stored in our database. Throughout the course, students revise their essays utilizing our platform through multiple iterations. The conversation and revision are based on what they learned from the course, as students give the first prompt indicating the contents covered in the class. Students also rate the helpfulness of every response from ChatGPT. After each session, their interim essays are saved for use in subsequent iterations. 

Unlike IRW and AW courses, SW students submit a paragraph of their choice to get help with while using the platform. This is because the course is designed to write a complete research paper over the semester by adding sections, and research papers cannot be completed during an in-class timed setting like essay writing. Therefore, data shown in Table \ref{tab:platform_data}a is an optional element in the platform for SW. Students with a paragraph can revise the essay and converse with ChatGPT, similar to IRW and AW user flow. Whereas students who do not submit any paragraph to edit can still review the course content and have conversations with ChatGPT, a personalized English writing teacher, to deepen their understanding.

We illustrate actual interaction data collected from \texttt{RECIPE} with students taking IRW, AW, and SW. A selected portion of the interactive dialogue data is described in Table \ref{tab:irw_dialogue} and \ref{tab:sw_dialogue} of Appendix \ref{appendix:interaction_data}. As shown in Table \ref{tab:sw_dialogue}, the student followed the user scenario as we intended. The student described the content of the class in the first prompt, which is focused on methodology, and submitted the methodology section of his or her own paper. Furthermore, after receiving ChatGPT's response suggesting three possible improvements to the paper, the student incorporated one of the suggestions into his or her own writing. In addition, a student taking IRW asked, ``Is 7 and 8 grammatical error?'' after ChatGPT listed up a revised version with ten grammatical errors. The student continues to question the response from ChatGPT by switching to Korean from English. Both students did not simply accept all the suggestions but sought clarification on suggested revisions that they did not understand. Therefore, this suggests that students are engaged in using \texttt{RECIPE}, and perform critical thinking, indicating the potential benefit of \texttt{RECIPE} for learning effect.

\section{Interview for Further Development}

We conducted two interviews to further scrutinize the needs for future work on \texttt{RECIPE}: (1) a focus group interview (FGI) among six students who had already taken at least one EFL writing course and (2) a 1:1 in-depth interview with an instructor who had taught EFL courses for 6 years, including IRW and AW.
Three students (S1, S2, S3) had taken undergraduate courses (IRW and AW), and the other three students (S4, S5, S6) had taken a graduate course (SW).
Detailed educational backgrounds of six student interviewees are depicted in Table \ref{tab:interviewee} Appendix \ref{appendix:interview}. We asked both students and instructor to use \texttt{RECIPE} and suggest future directions to ChatGPT-integrated educational platforms. In general, both of them acknowledged the usefulness of ChatGPT in EFL writing and expressed the need for a specialized platform integrated with ChatGPT in the EFL writing courses.

\paragraph{Recommendation for optimal prompts}
Although we provided instructions on the initial prompt and an example for the writing exercise, S5 and S6 reported difficulties in initiating a dialogue with ChatGPT.
S1 mentioned that sharing prompts that had effectively worked in the focus group was helpful.
Both S1 and S4 recommended implementing a tool to share the learning history of each student, including successful prompts and dialogue pattern logs.
We believe that the assessment data on each student's prompt and ChatGPT's response collected through this platform can be utilized to enhance \texttt{RECIPE}'s prompt engineering techniques and build a recommendation model for optimal prompts.

\paragraph{Personalized persona setup}
The instructor noted the need for personalized responses from ChatGPT based on students' situations, such as English proficiency level.
Additionally, S1 observed that a detailed persona setup with diverse perspectives, such as EFL professors, domain experts, and slightly advanced level EFL learners, made ChatGPT generate insightful critiques.
\texttt{RECIPE}'s persona, acting as an instructor of EFL writing courses, can be broadened in a diverse way so that students may choose the persona that best suits their purpose.

\paragraph{Student-initiative platform design}
The instructor remarked that continuous trials and errors in interactions with ChatGPT about revising their essays would be essential for their language learning.
The instructor also commented that those interactions led by students' initiative would facilitate their writing skills.
In particular, S1 suggested generating errorful essays relevant to what students learned through ChatGPT as learning materials and asking them to revise those essays by themselves, step by step, as ChatGPT guided them.

\paragraph{Automated essay scoring}
The instructor emphasized the need for automatically-measured essay scores and explanations for evaluation based on their own rubrics or grading systems.
While some state-of-the-art automated essay scoring (AES) systems for overall scores exist, it is nearly inaccessible for EFL instructors who don't have much background in AI techniques to incorporate and fine-tune them based on their own rubrics.
We believe that instructor-friendly platforms based on interactions with ChatGPT may mitigate this issue.

\paragraph{Refined user interface}
S5 suggested that a better user interface between two windows on the writing exercise of \texttt{RECIPE} would increase engagement among EFL students.
Specifically, text visualization or toggles for comparison of a student's original essay and ChatGPT's suggestions may be useful.

\section{Conclusion}
In this paper, we investigated students’ perception and usage of generative AI, including ChatGPT, in academic courses. Our results indicated that most students reported positive experiences with ChatGPT for general, academic, and essay-writing purposes. However, students with a limited understanding of LLMs faced challenges as expecting ChatGPT to find ground sources for their writings, and this hallucination is one of the major limitations of generative AI. To address this issue, we introduce \texttt{RECIPE}, a novel language learning platform, that leverages ChatGPT to cater to students’ needs and enhance their English writing skills with two types of prompts. We aim to collect interaction data throughout the semester on students' perception and usage of ChatGPT in English writing through \texttt{RECIPE}. The interaction data we gather with \texttt{RECIPE} has the potential to serve as a baseline for developing an effective paradigm for integrating ChatGPT in academic settings.

\begin{acks}
  This work was supported by Elice.  
\end{acks}

\bibliographystyle{ACM-Reference-Format}
\bibliography{sample-base}

\clearpage
\appendix
\begin{appendices}
\section{Details on Preliminary Questionnaire}\label{appendix:preliminary_study}

\subsection{Study Design}
This section describes how we design preliminary questionnaires.
Students are asked to answer 5-scale Likert questions in Table \ref{tab:preliminary_study_likert_question} towards their perception and usage of ChatGPT.
We also provide \textsc{Others} option for multi-select questions to get free-form inputs from students.

\begin{table}[htb!]
    \caption{5-scale Likert questions asked to study participants in the preliminary questionnaire.}
    \label{tab:preliminary_study_likert_question}
    \centering
    \begin{tabularx}{\columnwidth}{@{}>{\raggedright}p{0.27\columnwidth}|>{}X@{}}
    \toprule
    \multicolumn{2}{c}{Student's expectation on their EFL writing course}                                                                         \\ \midrule
    \textsc{motivation}    & How motivated are you to learn new things through this class?                                                                 \\
    \textsc{engagement}    & How willing are you to actively engage in this class?                                                                         \\
    \textsc{confidence}    & How confident are you in your ability to succeed in this class?                                                               \\ \midrule
    \multicolumn{2}{c}{Student's understanding of generative AI}                                                                                  \\ \midrule
    \textsc{LLM understanding} & How much do you understand about how Generative AI including ChatGPT works? (e.g., dataset, training method, LLM, limitation) \\ \midrule
    \multicolumn{2}{c}{Student's experience with ChatGPT}                                                                                         \\ \midrule
    \textsc{perceived helpfulness}          & How helpful was ChatGPT?                                                                                                      \\
    \textsc{trustworthiness}         & How much did you trust the response of ChatGPT?                                                                               \\
    \textsc{credibility}   & How credible was the information provided by ChatGPT?                                                                         \\
    \textsc{appropriateness of style/tone}         & How appropriate was the style or tone of responses provided by ChatGPT?                                                       \\
    \textsc{performance}       & How was the overall response quality of ChatGPT?                                                                              \\
    \textsc{overall satisfaction}  & How was the overall satisfaction using ChatGPT?                                                                               \\
    \textsc{referral intention} & How likely would you encourage your friends to use ChatGPT?                                                                   \\ \midrule
    \multicolumn{2}{c}{Student's expectation of using ChatGPT in courses}                                                                \\ \midrule
    \textsc{performance}       & How do you expect overall response quality using ChatGPT in this class?                                                       \\
    \textsc{credibility}   & How do you expect the credibility of information from ChatGPT regarding this class?                                           \\
    \textsc{overall satisfaction}  & How do you expect your overall satisfaction using ChatGPT in this class?                                                      \\ \bottomrule
    \end{tabularx}
\end{table}

\subsection{Findings}
This section describes detailed purposes of students' experience and expectations of using ChatGPT.
Students are asked to answer multi-select questions in Table \ref{tab:experience_general_purpose}, \ref{tab:expectation_writing_purpose}, and \ref{tab:expectation_academic_purpose}.
Numbers in each cell indicate the percentage of students who replied as the purpose and the parenthesis indicate the rank, respectively.

\begin{table}[htb!]
    \caption{Multi-select questions for the general purpose of ChatGPT use asked to study participants in the preliminary questionnaire. Please choose the purpose(s) for using ChatGPT.}
    \label{tab:experience_general_purpose}
    \centering
    \begin{tabularx}{\columnwidth}{@{}p{0.6\columnwidth}|XX@{}}
    \toprule
    Purpose                                                                                     & \multicolumn{1}{c|}{Undergrad.}             & \multicolumn{1}{c}{Grad.}                 \\ \midrule
    For fun                                                                                     & \multicolumn{1}{r|}{45.95 (2)}              & \multicolumn{1}{r}{50.82 (1)}             \\
    To get help with academic work in college (e.g., assignments, quiz, test)                   & \multicolumn{1}{r|}{51.65 (1)}              & \multicolumn{1}{r}{42.62 (3)}             \\
    To improve English skills (e.g, reading, writing, grammar, vocabulary, listening, speaking) & \multicolumn{1}{r|}{4.60 (7)}               & \multicolumn{1}{r}{21.31 (7)}             \\
    To translate texts                                                                          & \multicolumn{1}{r|}{9.89 (5)}              & \multicolumn{1}{r}{27.87 (5)}             \\
    To get help with research                                                                   & \multicolumn{1}{r|}{4.49 (7)}               & \multicolumn{1}{r}{48.36 (2)}             \\
    To assist coding and handling error messages                                                & \multicolumn{1}{r|}{13.19 (4)}              & \multicolumn{1}{r}{22.95 (6)}             \\
    To learn more about AI and chatbot                                                          & \multicolumn{1}{r|}{21.98 (3)}              & \multicolumn{1}{r}{30.33 (4)}             \\
    To get advice on personal problems or issues                                                & \multicolumn{1}{r|}{6.59 (6)}               & \multicolumn{1}{r}{7.38 (8)}              \\ \bottomrule
    \end{tabularx}
\end{table}

\begin{table}[htb!]
    \caption{Multi-select questions for English writing purpose of ChatGPT use asked to study participants in the preliminary questionnaire. By using ChatGPT in writing class, what kind of help do you expect from ChatGPT?}
    \label{tab:expectation_writing_purpose}
    \centering
    \begin{tabularx}{\columnwidth}{@{}p{0.6\columnwidth}|XX@{}}
    \toprule
    Purpose                                                             & \multicolumn{1}{c|}{Undergrad.} & \multicolumn{1}{c}{Grad.} \\ \midrule
    To translate texts                                                  & \multicolumn{1}{r|}{40.66 (5)}  & 51.64 (5)                 \\
    To get help with assignments                                        & \multicolumn{1}{r|}{34.07 (7)}  & 25.41 (8)                 \\
    To review the lecture                                               & \multicolumn{1}{r|}{12.09 (9)}  & 13.93 (10)                \\
    To brainstorm ideas                                                 & \multicolumn{1}{r|}{43.96 (4)}  & 34.43 (5)                 \\
    To find sources to support my writing                               & \multicolumn{1}{r|}{53.85 (2)}  & 32.79 (6)                 \\
    To get assessed and receive feedback                                & \multicolumn{1}{r|}{48.35 (3)}  & 45.08 (3)                 \\
    To get help with punctuation, spelling, capitalization, and grammar & \multicolumn{1}{r|}{56.04 (1)}  & 61.48 (1)                 \\
    To ask questions about the lecture                                  & \multicolumn{1}{r|}{7.69 (10)}  & 15.57 (9)                 \\
    To generate a sample essay or paragraph                             & \multicolumn{1}{r|}{30.77 (8)}  & 31.15 (7)                 \\
    To revise a draft                                                   & \multicolumn{1}{r|}{40.66 (5)}  & 42.62 (4)                 \\ \bottomrule
    \end{tabularx}
\end{table}

\begin{table*}[htb!]
    \caption{Multi-select questions for the academic purpose of ChatGPT use asked to study participants in the preliminary questionnaire. (a) If you have used ChatGPT for your academic work in college, what kind of academic work(s) did you use ChatGPT for? (b) For what kind of academic work do you plan to use ChatGPT?}
    \label{tab:expectation_academic_purpose}
    \centering
    \begin{tabular}{@{}l|cc|cc@{}}
    \toprule
    \multirow{2}{*}{Purpose}                                & \multicolumn{2}{c|}{(a) experience}                                    & \multicolumn{2}{c}{(b) expectation}                             \\ \cmidrule(l){2-5} 
                                                            & \multicolumn{1}{c|}{Undergrad.}       & \multicolumn{1}{c|}{Grad.}     & \multicolumn{1}{c|}{Undergrad.} & \multicolumn{1}{r}{Grad.}     \\ \midrule
    To translate texts                                      & \multicolumn{1}{r|}{16.48 (3)}        & \multicolumn{1}{r|}{31.97 (2)} & \multicolumn{1}{r|}{50.60 (2)}  & \multicolumn{1}{r}{61.86 (1)} \\
    To get help with assignments                            & \multicolumn{1}{r|}{38.46 (1)}        & \multicolumn{1}{r|}{36.89 (1)} & \multicolumn{1}{r|}{73.49 (1)}  & \multicolumn{1}{r}{59.79 (2)} \\
    To get answers for quizzes or tests                     & \multicolumn{1}{r|}{9.89 (6)}        & \multicolumn{1}{r|}{5.74 (6)}  & \multicolumn{1}{r|}{25.30 (6)}  &\multicolumn{1}{r}{11.34 (7)}  \\
    To prepare for quizzes or tests                         & \multicolumn{1}{r|}{15.38 (4)}        & \multicolumn{1}{r|}{13.11 (5)} & \multicolumn{1}{r|}{38.55 (4)}  & \multicolumn{1}{r}{20.62 (6)} \\
    To review the lectures                                  & \multicolumn{1}{r|}{15.38 (4)}        & \multicolumn{1}{r|}{4.92 (7)}  & \multicolumn{1}{r|}{34.94 (5)}  & \multicolumn{1}{r}{21.65 (5)} \\
    To ask questions about the lecture                      & \multicolumn{1}{r|}{20.88 (2)}        & \multicolumn{1}{r|}{14.75 (4)} & \multicolumn{1}{r|}{46.99 (3)}  & \multicolumn{1}{r}{34.02 (4)} \\
    To learn more in depth (beyond what is taught in class) & \multicolumn{1}{r|}{7.69 (7)}         & \multicolumn{1}{r|}{15.57 (3)} & \multicolumn{1}{r|}{22.89 (7)}  & \multicolumn{1}{r}{36.08 (3)} \\ \bottomrule
    \end{tabular}
\end{table*}

\section{Details on Interview}\label{appendix:interview}

The educational backgrounds of six student interviewees in FGI, including education level, EFL course(s) taken, and frequency and purpose of ChatGPT use, are described in Table \ref{tab:interviewee}.

\begin{table*}[hbt!]
    \caption{Educational Backgrounds of Students}
    \label{tab:interviewee}
    \centering
    \begin{tabular}{@{}ccccl@{}}
    \toprule
    Student & Current Program    & Class Taken  & Frequency of ChatGPT Use & Purpose of ChatGPT Use                                                                                   \\ \midrule
    S1      & Masters   & IRW, AW      & Often                    & \begin{tabular}[c]{@{}l@{}}\tabitem To translate texts\\ \tabitem To paraphrase texts\end{tabular}       \\ \midrule
    S2      & Bachelors & IRW, AW      & Rarely                   & \tabitem For fun                                                                                         \\ \midrule
    S3      & Doctoral  & IRW, AW      & Occasionally             & \tabitem To get help with research                                                                       \\ \midrule

    S4      & Doctoral  & SW           & Always                   & \begin{tabular}[c]{@{}l@{}}\tabitem To translate texts\\ \tabitem To get help with research\end{tabular} \\ \midrule
    S5      & Doctoral  & SW           & Occasionally             & \tabitem To get help with research                                                                       \\ \midrule
    S6      & Masters   & SW           & Often                    & \tabitem To assist coding                                                                                \\ \bottomrule
    \end{tabular}
\end{table*}

\newpage
\section{Interaction Data from \texttt{RECIPE}}\label{appendix:interaction_data}
Table \ref{tab:irw_dialogue} and \ref{tab:sw_dialogue} reveals th detailed interactions and dialogues illustrated in \S\ref{sec:user_scenario}.
Students ask the initial prompt to ChatGPT in \texttt{RECIPE}, and they are asked to assess ChatGPT's response on a 5-scale regarding the helpfulness of each response from ChatGPT.
They can edit their essay after each dialogue turn and continue the conversation after essay revision based on suggestions from ChatGPT.

\begin{table*}[htb!]
    \caption{Interactive dialogue data collected from \texttt{RECIPE} in Intermediate English Reading \& Writing (undergraduate course)}
    \label{tab:irw_dialogue}
    \centering
    \begin{tabularx}{\textwidth}{@{}X|X|c|c@{}}
    \toprule
    \multicolumn{1}{c|}{Student's Prompt}                                                                                       & \multicolumn{1}{c}{ChatGPT's Response}                                                                                                                                                                 & Helpfulness & Essay Edits \\ \midrule
    {\begin{tabularx}{\linewidth}[t]{@{}X@{}}<\textit{Student's original essay}>\\ Is there any grammatical errors in my essay?\end{tabularx}} & {\begin{tabularx}{\linewidth}[t]{@{}X@{}}Yes, there are a few grammar errors in your essay. Let's go over some of them and how to revise them:\\ \\ <\textit{Ten grammatical error feedbacks in order}>\end{tabularx}} & 5           & \ding{52}   \\ \midrule
    Is 7 and 8 grammatical error?                                                                                              & {\begin{tabularx}{\linewidth}[t]{@{}X@{}}Yes, 7 and 8 are grammatical errors. Let me explain why:\\ \\ <\textit{Explanations for grammatical error 7 and 8}>\end{tabularx}}                                            & 2           &             \\ \midrule
    {\begin{tabularx}{\linewidth}[t]{@{}X@{}}7번 문장에서 other말고 다른 부분은 문법적 오류가 아니지 않아?\\ \textit{(Isn't the part other than ``other'' in sentence 7 a grammatical error?)}\end{tabularx}} & {\begin{tabularx}{\linewidth}[t]{@{}X@{}}네, 다른 부분들은 문법적 오류가 아니라 문체나 표현에 있어서 다소 어색하거나 효율적이지 않은 것으로 교정해야 할 때도 있습니다. 다른 부분들도 함께 교정해보시면 더욱 발전된 글쓰기를 할 수 있을 것입니다.\\ \textit{(No, you might need to correct other parts that are not grammatical errors but somewhat awkward or inefficient in style or expression. If you correct those parts as well, you can write a better essay.)}\end{tabularx}} & 2           &             \\ \bottomrule

    \end{tabularx}
\end{table*}

\begin{table*}[htb!]
    \caption{Interactive dialogue data collected from \texttt{RECIPE} in Scientific Writing (graduate course)}
    \label{tab:sw_dialogue}
    \centering
    \begin{tabularx}{\textwidth}{@{}X|X|c|c@{}}
    \toprule
    \multicolumn{1}{c|}{Student's Prompt}                                                                                                                                                                                     & \multicolumn{1}{c|}{ChatGPT's Response}                                                                                                                                                                                                                                                                                                                                                     & Helpfulness            & Essay Edits \\ \midrule
    I learned about the basic composition of the methodology part. Normally, methodology part contains information of material, method, procedure of the experiment.                                                          & Yes, that is correct! The methodology section of a scientific paper is where you describe the materials, methods, and procedures used to conduct your research. It's important to be clear and concise so that others can understand your research and potentially replicate it. Did you have any specific questions about the methodology section or any other part of scientific writing? & \multicolumn{1}{c|}{4} & \ding{52}   \\ \midrule
    I wrote the methodology of my study. Could you have a look and give me some feedback?                                                                                                                                     & Yes, I would be happy to take a look at your methodology section and provide you with feedback. Please share your work with me, and I'll do my best to help you improve it.                                                                                                                                                                                                                 & \multicolumn{1}{c|}{4} &             \\ \midrule
    <\textit{Methodology section from student's research paper}>                                                                                                                                                              & {\begin{tabularx}{\linewidth}[t]{@{}X@{}}Great job on writing your methodology section! I have a few suggestions that could help you improve it even more. \\ \\ <\textit{Three suggestions on writing methodology section in order}>\\ ...\end{tabularx}}                                                                                                                                                  & \multicolumn{1}{c|}{3} & \ding{52}   \\ \midrule
    If I add following sentence at the beginning of the paragraph, would it be enough to explain the context? ``<\textit{A new sentence written by the student based on the first suggestion from previous ChatGPT response}>'' & Yes, adding that sentence would help provide context for your study and give readers a better understanding of the rationale behind the methodology you used. It sets up the research question and hypothesis well, and makes it clear why the study was designed the way it was.                                                                                                           & \multicolumn{1}{c|}{4} & \ding{52}   \\ \bottomrule
    \end{tabularx}
\end{table*}
\end{appendices}

\end{document}